\documentclass[aps,prl,reprint,superscriptaddress,nofootinbib]{revtex4-2}

\usepackage[utf8]{inputenc}
\usepackage[T1]{fontenc}
\usepackage{lmodern}

\usepackage{graphicx}
\usepackage{amsmath,amssymb,bm}
\usepackage{siunitx}
\usepackage{hyperref}
\usepackage{braket}
\usepackage{physics}
\usepackage{xcolor}
\usepackage{bibunits}
\usepackage{orcidlink}

\hypersetup{
  colorlinks=true,
  linkcolor=blue,
  citecolor=blue,
  urlcolor=blue
}

\begin{document}


\title{High-flux sub-Poissonian twin fields generation from warm atomic vapor}

\author{Priya Drashni\orcidlink{0009-0009-5676-6033}}
\email{pd26h@fsu.edu}
\affiliation{Department of Electrical \& Computer Engineering, FAMU–FSU College of Engineering, Florida State University, Tallahassee, FL 32306, USA}

\author{Hari P.~Lamsal}
\affiliation{Department of Physics \& Astronomy, University of Tennessee, Chattanooga, TN 37403, USA}

\author{Belle A.~White}
\affiliation{Department of Physics \& Astronomy, University of Tennessee, Chattanooga, TN 37403, USA}

\author{Noah A.~Crum}
\affiliation{Department of Physics \& Astronomy, University of Tennessee, Knoxville, TN 37996, USA}

\author{George Siopsis\orcidlink{0000-0002-1466-2772}}
\affiliation{Department of Physics \& Astronomy, University of Tennessee, Knoxville, TN 37996, USA}

\author{Tian Li\orcidlink{0000-0003-2993-0386}}
\email{tli6@fsu.edu}
\affiliation{Department of Electrical \& Computer Engineering, FAMU–FSU College of Engineering, Florida State University, Tallahassee, FL 32306, USA}


\begin{abstract}

 We demonstrate the generation of sub-Poissonian twin fields via near-degenerate spontaneous four-wave mixing (SFWM) in warm $^{85}\mathrm{Rb}$ vapor at 795~nm. When seeded with a weak coherent field, the generated twin beams exhibit approximately $5.5~\mathrm{dB}$ of intensity-difference squeezing in free space and retain about $3~\mathrm{dB}$ after coupling into polarization-maintaining (PM) fibers. Under vacuum seeding, time-resolved photon-counting measurements yield Mandel parameters of $Q\approx-0.7$ for each individual field, demonstrating strong photon-number squeezing. To explain these observations, we develop a finite-resource saturation model in which occupation-dependent SFWM gain, arising from competition for a finite nonlinear gain resource, suppresses large photon-number fluctuations within an effective collective mode selected by the PM-fiber spatial projection, thereby producing the observed negative Mandel-$Q$ parameters. The temporal cross-correlation between the twin photons exhibits a distinctive flat-topped profile resulting from the interplay of multiple $\chi^{(3)}$ processes in the atomic medium and is in excellent agreement with the theoretical model. Combining high photon flux, near-resonant operation, robust sub-Poissonian photon statistics, and fiber compatibility, this source provides a promising platform for scalable quantum-enhanced sensing and quantum information processing.

\end{abstract}

\maketitle

\textit{Introduction –}
Nonclassical states of light with reduced photon number fluctuations are essential resources for quantum-enhanced sensing and precision measurements~\cite{Giovannetti2011,Lawrie2019}. In particular, sub-Poissonian photon statistics, characterized by fluctuations suppressed below the Poissonian level, constitute a direct signature of nonclassicality~\cite{Mandel1979,ShortMandel1983}. Using photon counting measurements, such states can surpass the classical shot-noise limit (SNL), enabling enhanced sensitivity with applications in quantum imaging, optical metrology, and quantum information processing~\cite{Brida2010,Pirandola2018,Degen2017,RuoBerchera2019}.

Bright-seeded four-wave mixing (BFWM) and spontaneous four-wave mixing (SFWM) in warm atomic vapors are among the most versatile $\chi^{(3)}$ platforms for generating nonclassical states of light at atomic transition wavelengths. In these configurations, the FWM process produces spectrally narrowband, strongly correlated photon pairs, commonly referred to as Stokes and anti-Stokes photons in SFWM, and probe and conjugate beams in BFWM, whose correlations originate from ground-state coherence in the atomic medium. These systems have been widely used to generate heralded anti-Stokes photons with \textit{conditional} sub-Poissonian statistics, typically realized under heralded detection schemes at relatively low photon flux~\cite{Shu2016,Lamperti2014,Du2008,Huang2025}, as well as two-mode quadrature and intensity-difference squeezing~\cite{McCormick2007,deAraujo2024,Glorieux2011,Li2017PSA}. \textcolor{black}{However, the photon-number statistics of individual fields in the single spatio-temporal detection mode regime, quantified, for example, by the Mandel-$Q$ parameter, remain largely unexplored in warm-vapor $\chi^{(3)}$ platforms.}

In parametric $\chi^{(3)}$ processes, the generated correlated fields typically exhibit super-Poissonian photon-number fluctuations associated with thermal-like photon statistics~\cite{Boyd2015NonlinearOptics,MandelWolf1995}. This behavior arises because tracing over the correlated partner yields a mixed thermal state for each field, while detection typically averages incoherently over many independent spatio-temporal modes~\cite{MandelWolf1995}. In warm-vapor-based $\chi^{(3)}$ media, the finite interaction volume intrinsically supports multiple spatial and spectral modes, as evidenced by spatially multimode twin-beam generation in four-wave mixing~\cite{Boyer2008}, so the observed photon-number distribution often resembles a thermal average~\cite{Boyd2015NonlinearOptics}. Because directly resolving the mode structure of individual fields in warm atomic vapor is experimentally challenging, many prior studies have instead relied on number-squeezing techniques, such as two-mode intensity-difference squeezing measured with photodiodes, to infer underlying quantum correlations~\cite{Prajapati2021}. \textcolor{black}{Demonstrating access to near single spatio-temporal detection modes in a highly multimode warm-vapor system, and thereby revealing strong sub-Poissonian photon-number statistics at high photon flux, constitute the central objectives of this work.}

In this work, we investigate the quantum statistical properties of twin fields generated via near-degenerate FWM in warm $^{85}$Rb vapor, {characterized in a near single spatio-temporal detection mode regime.} When seeded with a weak coherent input field, intensity-difference squeezing provides a direct measure of the photon-number correlations between the twin beams. Under vacuum seeding, arrival-time-resolved coincidence detection instead provides direct insight into the temporal correlations between the output twin fields arising from the underlying $\chi^{(3)}$ nonlinear interaction. By combining the arrival-time-resolved  measurement with spatial-mode filtering using polarization-maintaining (PM) single-mode fiber, {we directly probe the photon-number statistics of each field within the selected near-single spatio-temporal detection mode.} We observe strong sub-Poissonian statistics in each field individually, confirmed by the negativity of their respective Mandel-$Q$ parameters, while the pair collectively exhibits distinct temporal correlations governed by multiple FWM processes in the atomic medium. With photon fluxes in the MHz range, our approach enables high-flux, sub-Poissonian twin fields generation from warm atomic vapor in a fiber-compatible platform, paving the way for scalable quantum sensing applications.

\begin{figure}[t]
\centering
\includegraphics[width=1\linewidth, trim=0 120 5 60, clip]{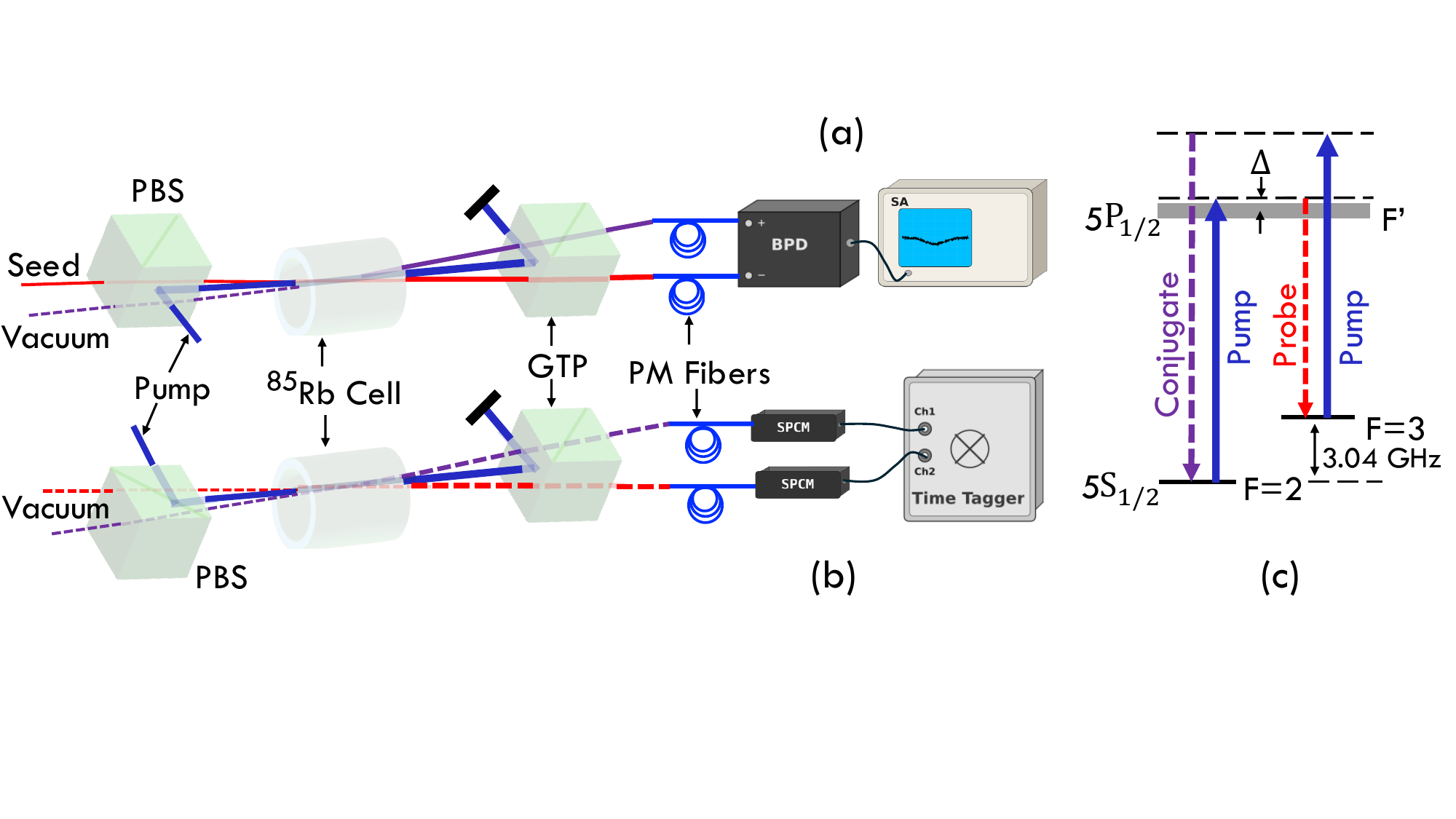}
\caption{
Experimental setups and energy-level diagram for FWM processes in warm $^{85}\mathrm{Rb}$ vapor.
(a) Bright-seeded FWM (BFWM): A strong coherent pump beam and a weak coherent seed beam intersect within the vapor cell. The resulting twin beams, i.e., the probe (red solid line) and conjugate (purple dashed line) beams, are coupled into PM fibers and detected with a BPD. The photocurrent difference is analyzed using a radio-frequency SA to quantify intensity-difference squeezing. 
(b) Spontaneous FWM (SFWM): Both probe and conjugate modes are unseeded. The generated photon pairs are collected into the same PM fibers that satisfy the same phase-matching condition established in (a). The bottom dotted red line corresponds to the probe, and the dotted purple line corresponds to the conjugate. The arrival times of correlated photons are recorded by two SPCMs and processed using a coincidence measurement similar to TCSPC, but with each individual photon's arrival time explicitly time-tagged for Mandel-$Q$ parameter analysis. 
(c) Double-$\Lambda$ energy-level diagram illustrating the single-photon detuning $\Delta$ of the pump field and the probe frequency offset corresponding to the ground-state hyperfine splitting of $^{85}\mathrm{Rb}$. The conjugate field is generated via the $\chi^{(3)}$ nonlinear process when both energy conservation and phase-matching conditions are satisfied in the atomic vapor.
}
\label{fig:experiment}
\end{figure}

\textit{Experimental scheme –}
Our experimental scheme is based on a double-$\Lambda$ FWM configuration in warm $^{85}$Rb vapor. This platform generates quantum-correlated probe and conjugate beams in the BFWM regime, as well as probe and conjugate photon pairs in the SFWM regime, as illustrated in Figs.~\ref{fig:experiment}(a) and \ref{fig:experiment}(b), respectively. In the BFWM configuration, a strong pump beam intersects with a cross-polarized weak coherent seed within the vapor cell. The generated probe and conjugate twin beams, indicated by the solid red and purple paths, are separated from the pump using a Glan–Taylor polarizer (GTP) after the cell. They are then coupled into two PM fibers and detected with a balanced photodetector (BPD). The resulting photocurrent difference is analyzed using a radio-frequency spectrum analyzer (SA) to quantify intensity-difference squeezing. In the SFWM configuration, both probe and conjugate modes are unseeded, and the generated photon pairs are collected into the same PM fibers as in Fig.~\ref{fig:experiment}(a), thereby satisfying the same phase-matching condition. After fiber coupling, the correlated photons are directed to two single-photon counting modules (SPCMs) and processed by a time tagger for coincidence analysis, which is similar to time-correlated single-photon counting (TCSPC), but with each individual photon's arrival time explicitly time-tagged for Mandel-$Q$ parameters analysis. The double-$\Lambda$ energy-level diagram is shown in Fig.~\ref{fig:experiment}(c), where the pump beam is detuned by $\Delta = 800~\mathrm{MHz}$ to the blue of the $\lvert 5S_{1/2}, F=2 \rangle \rightarrow \lvert 5P_{1/2}, F' \rangle$ transition.

In our experiment, an external-cavity diode laser operating near the 795~nm D$_1$ line provides both the pump and the frequency-shifted seed beam, as shown in Fig.~\ref{fig:experiment}(a). The pump, with power ranging from 5 to 450~mW and a $1/e^2$ beam waist of 1~mm at the center of the cell, drives the FWM process. The weak coherent seed is frequency down-shifted by 3.04~GHz from the pump using an acousto-optic modulator (AOM) to match the ground-state hyperfine splitting, with an input power of 10–100~$\mu$W and a $1/e^2$ beam waist of 0.5~mm at the center of the cell. The pump and seed beams, prepared with orthogonal linear polarizations, intersect at an angle of 0.3$^\circ$ within a 12.5~mm-long isotopically enriched $^{85}$Rb vapor cell maintained at 92~$^\circ$C. Through the $\chi^{(3)}$ interaction, two pump photons are converted into either the seeded mode (BFWM) or vacuum modes (SFWM), generating an amplified probe beam (or amplified vacuum fluctuations) along with a correlated and frequency-shifted conjugate beam (or correlated amplified vacuum fluctuations) at the phase-matched angle. To maintain high fluxes for the probe and conjugate photons, only multiple irises are used after the GTP to suppress residual pump light and forward-scattered fluorescence. \textit{No narrowband etalon or Fabry–Perot cavity filtering is employed}, in contrast to the scheme demonstrated in~\cite{Shu2016}. The resulting probe and conjugate fields are then coupled into two PM fibers with coupling efficiencies of 89~\% and 87~\%, respectively, providing effective spatial filtering for single spatio-mode selection.

\begin{figure}[t]
\centering
\includegraphics[width=1\linewidth]{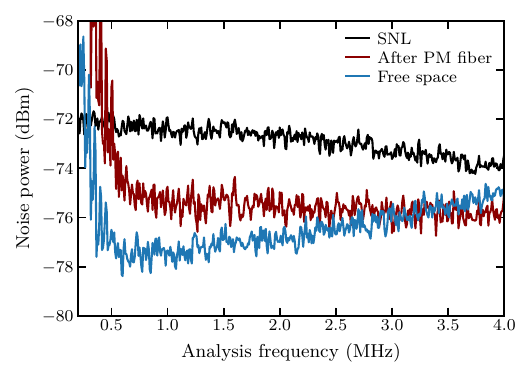}
\caption{Intensity-difference squeezing of twin beams under free-space and PM fiber–coupled conditions, measured using a SA. The noise power of the probe–conjugate photocurrent difference, normalized to the SNL (black curve), is shown as a function of analysis frequency of the SA. The free-space measurement (blue curve) exhibits up to 5.5~dB of noise suppression below the SNL. After coupling both beams into PM fibers (red curve), up to 3~dB of squeezing is preserved. The resolution bandwidth and video bandwidth are 10~kHz and 300~Hz, respectively.}
\label{fig:squeezing}
\end{figure}

\textit{Intensity-difference squeezing –} In the BFWM regime, nonclassical temporal correlations between the probe and conjugate beams are characterized via intensity-difference squeezing measurements~\cite{McCormick2007, deAraujo2024,Sim2025}. The shot-noise level is calibrated using a coherent beam with the same total optical power as the combined probe and conjugate beams, which in this experiment is 450~$\mu$W, and is shown as the black curve in Fig.~\ref{fig:squeezing}. In free space without PM fiber coupling, the probe–conjugate intensity difference exhibits up to 5.5~dB of squeezing below the shot-noise level (blue curve in Fig.~\ref{fig:squeezing}). Coupling both beams into PM fibers reduces the observed squeezing to close to 3~dB due to coupling losses and spatial filtering (red curve in Fig.~\ref{fig:squeezing}), while still demonstrating our experimental scheme's viability for fiber-based quantum platforms. \textcolor{black}{It is worth noting that the squeezing measured after the PM fiber coupling is calibrated according to the procedure outlined in Supplemental Note E.}

\begin{figure}[h]
\centering
\includegraphics[width=1\linewidth]{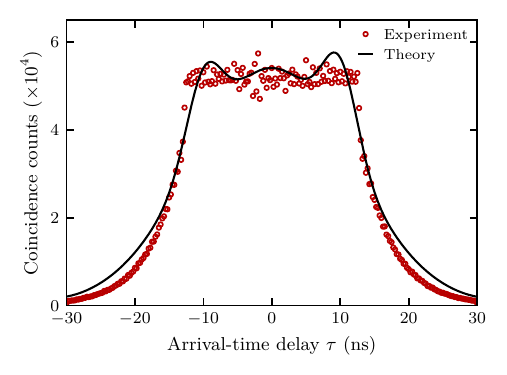}
\caption{Coincidence counts as a function of the arrival-time delay between the probe and conjugate photons (red open circles), shown together with the theoretical model prediction (black curve). The model incorporates the dominant SFWM process as well as two parasitic FWM contributions. Details about the theoretical prediction are provided in Supplemental Note A.}
\label{fig:model_fitting}
\end{figure}

\textit{Temporal correlation structure –} In the BFWM regime, the gain of the FWM process (5.0 in this experiment) results in a photon flux that exceeds the linear counting range of the SPCMs. We therefore study the temporal correlations between probe and conjugate photons in the SFWM regime, as shown in Fig.~\ref{fig:experiment}(b), where the lower photon flux enables time-resolved photon counting and coincidence analysis using SPCMs, a standard approach for characterizing biphoton emission dynamics~\cite{Willis2010,Liao2014,Shu2016}. In atomic media, the coincidence peak profile reflects the temporal coherence of the $\chi^{(3)}$ nonlinear interaction, including the effects of Doppler broadening in warm vapors~\cite{Shu2016}. In our experiment, the coincidence counts as a function of the relative time delay between the probe and conjugate photons are constructed from time-tagged detection events recorded on two channels of a time-tagger, as shown in Fig.~\ref{fig:experiment}(b). Raw timing data are parsed in Python to extract timestamp streams for individual beams, and coincidences at different arrival-time delays $\tau = t_{\mathrm{probe}} - t_{\mathrm{conjugate}}$ are counted from $-30~\si{\nano\second}$ to $+30~\si{\nano\second}$ using a unit time-bin width of \SI{250}{\pico\second}. The resulting coincidence temporal profile is shown in Fig.~\ref{fig:model_fitting}.

The measured coincidence temporal profile exhibits a broad, flat-topped central feature with slowly decaying wings extending over tens of nanoseconds. This behavior indicates that the probe–conjugate photon pairs are generated over a finite temporal window rather than at a sharply defined emission time, consistent with observations in other atomic biphoton sources where the coincidence temporal profile is governed by the medium’s phase-matching and coherence properties~\cite{Willis2010,Liao2014,Brecht2015}. The central plateau suggests an approximately uniform pair-generation probability over a range of arrival-time delays, enabled by the atomic $\chi^{(3)}$ nonlinearity. The extended wings reflect the persistence of temporal correlation associated with long-lived ground-state coherence.

To quantitatively describe this unique temporal behavior, we develop a theoretical model that accounts for the relevant emission processes in Doppler-broadened $^{85}\mathrm{Rb}$ vapor. The dominant SFWM process is described using a Doppler-averaged $\chi^{(3)}$ nonlinear susceptibility together with the longitudinal phase-matching condition. In addition, two parasitic FWM processes, each consisting of Raman scattering followed by spontaneous emission, are included to reproduce the flat-topped central feature. Details of the model are provided in Supplemental Note A. The model prediction, with parameters evaluated under realistic experimental conditions, is overlaid with the data in Fig.~\ref{fig:model_fitting}, showing excellent agreement.

\textit{Mandel-$Q$ parameter for individual fields –}
The photon-number fluctuations of the probe and conjugate photon fluxes are \textit{independently} characterized by the Mandel parameter $Q \equiv [\mathrm{Var}(n) - \langle n \rangle]/\langle n \rangle$, which quantifies deviations from Poissonian counting statistics. A coherent state yields $Q=0$, thermal light gives $Q>0$, and $Q<0$ indicates sub-Poissonian statistics, corresponding to suppressed photon-number fluctuations below the shot-noise level, with the limiting case $Q=-1$ corresponding to an ideal Fock state~\cite{MandelWolf1995}. The probe and conjugate photons generated via SFWM are analyzed by partitioning time-tagged detection events into temporal bins of width $\Delta t_{\mathrm{bin}}=\SI{100}{\nano\second}$. Photon-number statistics and Mandel-$Q$ parameters are then extracted independently for each field. The chosen bin width exceeds the SPCM dead time ($30~\mathrm{ns}$) while remaining below the mean inter-photon arrival time ($\sim 167~\mathrm{ns}$, corresponding to a photon flux of $6\times10^{6}~\mathrm{s^{-1}}$), ensuring that the measured statistics reflect the intrinsic quantum properties of the photon flux rather than detector related artifacts. 

The measured Mandel parameters for the probe and conjugate photon fluxes are $Q_{\mathrm{probe}} = -0.694 \pm 0.002$ and $Q_{\mathrm{conjugate}} = -0.732 \pm 0.003$, where the uncertainties denote one standard deviation obtained from $10^{4}$ iterations. These roughly $70~\%$ reductions in photon-number variance relative to Poissonian statistics demonstrate strong suppression of quantum noise below the shot-noise level \textit{for each field individually}. This behavior stands in marked contrast to prior demonstrations, where individual fields generated via nonlinear parametric processes typically exhibit super-Poissonian statistics ($Q>0$), reflecting thermal-like fluctuations~\cite{McCormick2007,Christ2011,Srivathsan2013}.

\textcolor{black}{We model the observed sub-Poissonian statistics as arising from finite-resource saturation of the SFWM process, in which the probability of generating an additional photon pair progressively decreases as the occupation of the effective collective mode preferentially selected by the PM-fiber spatial projection increases. Such occupation-dependent gain suppresses the high-photon-number tail relative to the thermal distribution expected for conventional undepleted SFWM, thereby narrowing the photon-number distribution. In the strongly occupied regime, the model predicts sub-Poissonian statistics when the effective unsaturated pair-generation rate exceeds approximately twice the removal rate. Notably, a generation-to-removal ratio of $r=5$ yields $Q\simeq-3/4$, consistent with the magnitude of the sub-Poissonian statistics observed experimentally. The relatively low pump power used in our experiment ($\sim 5~\mathrm{mW}$) may further favor operation outside the ideal undepleted-pump approximation, making finite-resource saturation and pump-mediated nonlinear feedback plausible under our experimental conditions. The detailed derivation of the model is provided in Supplemental Note B.}

Note that the Mandel-$Q$ parameters are obtained with a photon flux of approximately $6\times10^{6}~\mathrm{s^{-1}}$ per field, which is significantly higher than the heralded anti-Stokes photon rates reported in~\cite{Shu2016,Du2008,Zhao2014}. At low pump powers (below approximately $5~\mathrm{mW}$), the $\chi^{(3)}$ nonlinearity becomes too weak to support appreciable SFWM, and no sub-Poissonian photon statistics are observed. \textcolor{black}{One possible explanation is that, in this regime, the detected photon flux is increasingly dominated by fluorescence associated with linear absorption in the atomic vapor.} At higher pump powers, the increased probe and conjugate photon flux exceeds the linear response range of the SPCMs, and detector dead time inevitably introduces artificial antibunching into the measured statistics~\cite{Christ2011,Harder2016}. \textcolor{black}{It is worth noting that the Mandel-$Q$ parameter remains unchanged over  singles count rate ranging from $4.9\times10^{6}$ to $7.7\times10^{6} ~\mathrm{s^{-1}}$. The detailed flux-dependent measurements are included in Supplemental Note C.}

The observation of strongly sub-Poissonian statistics in both probe and conjugate photon fluxes, together with its robustness across a range of temporal bin widths (see Supplemental Note C), constitutes a central result of this work. It demonstrates that nonclassical suppression of photon-number fluctuations can be resolved in individual photon fluxes generated from a warm atomic-vapor $\chi^{(3)}$ system through appropriate spatial-mode filtering and time-resolved detection, despite the intrinsic multimode nature of the SFWM process.

\begin{table}[b]
\caption{
Individual-field Mandel-$Q$ parameters, evaluated with a bin width of $\Delta t_{\mathrm{bin}}=\SI{100}{\nano\second}$ \textcolor{black}{under the same singles count rate of $6\times10^{6}~\mathrm{s^{-1}}$ for different sources}. Uncertainties denote one standard deviation, obtained from $10^{4}$ iterations.}
\label{tab:mandel}
\centering
\begin{tabular*}{\columnwidth}{@{\extracolsep{\fill}}lcc}
\hline\hline
Source & $Q_{\mathrm{probe/signal}}$ & $Q_{\mathrm{conjugate/idler}}$ \\
\hline
Coherent & $0.000 \pm 0.001$ & $0.000 \pm 0.002$ \\
SPDC     & $+0.025 \pm 0.001$ & $+0.025 \pm 0.002$ \\
SFWM     & $-0.694 \pm 0.002$ & $-0.732 \pm 0.003$ \\
\hline\hline
\end{tabular*}
\end{table}

\begin{figure}[]
\centering
\includegraphics[width=1\linewidth]{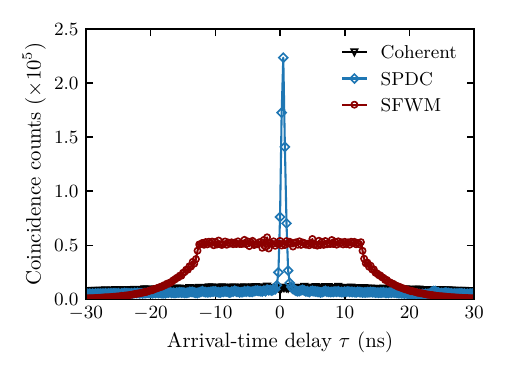}
\caption{Coincidence counts as a function of the arrival-time delay between photon pairs generated by coherent, SPDC, and SFWM sources \textcolor{black}{under the same singles count rate of $6\times10^{6}~\mathrm{s^{-1}}$}. The coherent source yields a flat, near-zero background (black curve). The SPDC source exhibits a narrow peak centered at near zero delay (blue curve), whereas the warm $^{85}\mathrm{Rb}$-vapor-based SFWM source displays a distinct flat-topped profile with slowly decaying wings (red curve).}
\label{fig:combined_plot}
\end{figure}

\textit{Comparison with other states of light –} To place these observations in context, we compare the temporal correlation structure of our warm $^{85}\mathrm{Rb}$-vapor-based SFWM source with that of a coherent source and a biphoton source (a commercial type-II SPDC  source) \textcolor{black}{under the same singles count rate of $6\times10^{6}~\mathrm{s^{-1}}$}. The results are shown in Fig.~\ref{fig:combined_plot}. The coherent source produces a near-zero, flat coincidence profile, reflecting the absence of temporal correlations, as expected. In contrast, the SPDC biphoton source exhibits a sharply peaked coincidence profile centered at near zero delay, indicating broadband biphoton generation with a short coherence time. This comparison highlights the distinct temporal structure arising from the $\chi^{(3)}$ nonlinear processes in warm $^{85}\mathrm{Rb}$ vapor, clearly differentiating it from both coherent light and SPDC-based biphoton sources~\cite{Jabir2017}.

For completeness, we also perform the Mandel-$Q$ parameter analysis for both the coherent and SPDC sources. The same binning and analysis procedure is applied to the two output fields obtained by splitting a coherent beam on a BS, as well as to the biphoton fields (i.e., signal and idler) generated by the SPDC source. The results are summarized in Table~\ref{tab:mandel}. The two coherent fluxes yield $Q \approx 0$ within statistical uncertainty, consistent with shot-noise-limited statistics. In contrast, the biphoton fields exhibit a small but statistically significant positive $Q$ (exceeding $20~\sigma$ from zero), consistent with the super-Poissonian marginal statistics expected for individual fields in SPDC processes.

\begin{figure}[t]
\centering
\includegraphics[width=1\linewidth]{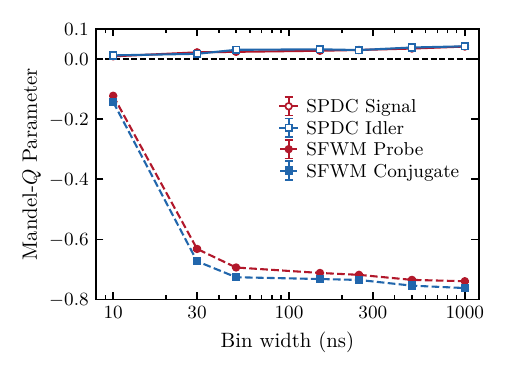}
\caption{\textcolor{black}{Measured Mandel-$Q$ parameters for different counting-bin widths for the SFWM and SPDC sources under the same singles count rates. \textit{The $x$-axis is displayed on a logarithmic} scale. The error bars represent one standard deviation obtained from $10^{4}$ iterations and are smaller than the graph symbols.}}
\label{fig:binwidth}
\end{figure}

\textcolor{black}{To verify that the observed negative Mandel-$Q$ parameters do not arise from detector related artifacts, additional measurements spanning a \textit{two-order-of-magnitude range} of counting-bin widths were performed using both the SPDC biphoton source and the SFWM source at the same detected singles count rates. As shown in Fig.~\ref{fig:binwidth}, the Mandel-$Q$ parameter for the SFWM source remains negative and largely unchanged over the entire two-order-of-magnitude range of counting-bin widths, whereas the corresponding $Q$ parameter for the SPDC source exhibits a slight positive trend with increasing bin width. At the shortest bin width of 10~ns, however, which is well below the detector dead time of 30~ns, the SFWM $Q$ parameter becomes less negative, indicating an increasing influence of detector-dead-time-induced artifacts. The individual Mandel-$Q$ parameters and their corresponding uncertainties are provided in Supplemental Note~C.
}

The simultaneous observation of pronounced sub-Poissonian statistics in both the probe and conjugate photon fluxes generated via the fiber-coupled, warm $^{85}\mathrm{Rb}$-vapor-based SFWM process demonstrates the suppression of photon-number fluctuations in each field \textit{without the need for heralding}, which was previously required to reveal sub-Poissonian behavior~\cite{Shu2016,Zhu2017}. These results highlight the coexistence of strong single-field photon-number squeezing along with twin-field-like temporal correlations.

\textcolor{black}{\textit{Conclusion –} We report pronounced sub-Poissonian statistics in both photon fluxes generated
via PM fiber-coupled, near-degenerate SFWM in warm $^{85}\mathrm{Rb}$ vapor,
characterized using time-resolved photon counting and Mandel-$Q$ analysis.
The probe and conjugate photon fluxes each exhibit approximately $70~\%$
suppression of photon-number variance below the SNL over a broad range of
temporal bin widths. These results demonstrate strong nonclassical behavior
at a high photon flux of approximately $6\times10^{6}~\mathrm{s^{-1}}$ per
field. To account for the observed negative Mandel-$Q$ parameters, we develop
a finite-resource saturation model in which the SFWM gain becomes
occupation dependent within an effective collective mode selected through
spatial-mode projection into the single-mode PM fiber. The resulting
suppression of higher-photon-number fluctuations provides a mechanism for
the emergence of sub-Poissonian statistics beyond the conventional
undepleted SFWM description. In the strongly occupied regime, the model
predicts negative Mandel-$Q$ parameters and yields $Q\simeq-3/4$ for an
effective generation-to-removal ratio of $r=5$, consistent with the magnitude
observed experimentally.}

\textcolor{black}{In addition, the temporal correlation between the probe and conjugate photon fluxes exhibits a distinctive flat-topped profile, in excellent agreement
with a theoretical model that incorporates both the dominant SFWM process
and two parasitic FWM contributions. These results indicate that PM fiber
coupling provides an effective spatial-mode projection that preferentially
selects a collective mode of the nonlinear SFWM dynamics. Combined with
time-resolved photon counting, this enables the observation of nonclassical
photon-number statistics in each individual field at high photon flux, in
contrast to prior warm-vapor-based SFWM studies relying on heralding at
relatively low photon flux~\cite{Shu2016,Zhu2017}.}

This platform provides a fiber-compatible source of spectrally narrowband nonclassical light at atomic transition wavelengths, while simultaneously enabling strong single-field sub-Poissonian statistics and coherence-limited temporal correlations. As such, it is well suited for integration into scalable quantum-enhanced sensing and information processing applications.

\textit{Acknowledgments –} BAW and TL acknowledge support from the U.S. National Science Foundation (NSF) through the ExpandQISE program under Award No. 2426699, and the NSF CCSS program under Award No. 2503630. HPL and TL also acknowledge support from the U.S. National Institute of Standards and Technology (NIST) through the CIPP program under Award No. 60NANB24D218. PD, NC, and GS acknowledge support from the U.S. Department of Energy, Office of Science, Office of Advanced Scientific Computing Research, through the Quantum Internet to Accelerate Scientific Discovery Program under Field Work Proposal 3ERKJ381. PD, NC, and GS also acknowledge support from the NSF under Award No. DGE-2152168.



\bibliographystyle{apsrev4-2}
\bibliography{references}
\noindent\rule{\linewidth}{0.4pt}


\clearpage
\onecolumngrid


\setcounter{NAT@ctr}{0}

\section*{Supplemental Material}

\renewcommand{\thefigure}{S\arabic{figure}}
\renewcommand{\thetable}{S\arabic{table}}
\renewcommand{\theequation}{S\arabic{equation}}

\setcounter{figure}{0}
\setcounter{table}{0}
\setcounter{equation}{0}

\makeatletter
\renewcommand{\fnum@figure}{Fig.~\thefigure}
\renewcommand{\fnum@table}{Table~\thetable}
\makeatother


\section*{Supplemental Note A: Theoretical model of the photon-arrival-time coincidence profile}

In this Supplemental Note, we present the theoretical model used to understand the coincidence temporal profile shown in the main text. The spontaneous four-wave mixing (SFWM) process occurs in a Doppler-broadened $^{85}$Rb vapor driven by a relatively strong pump field. In addition to the dominant SFWM process, we include two parasitic FWM processes comprised of Raman scattering followed by spontaneous emission.

\begin{figure}[h]
    \centering
    \includegraphics[width=\linewidth, trim=90 120 90 90, clip]{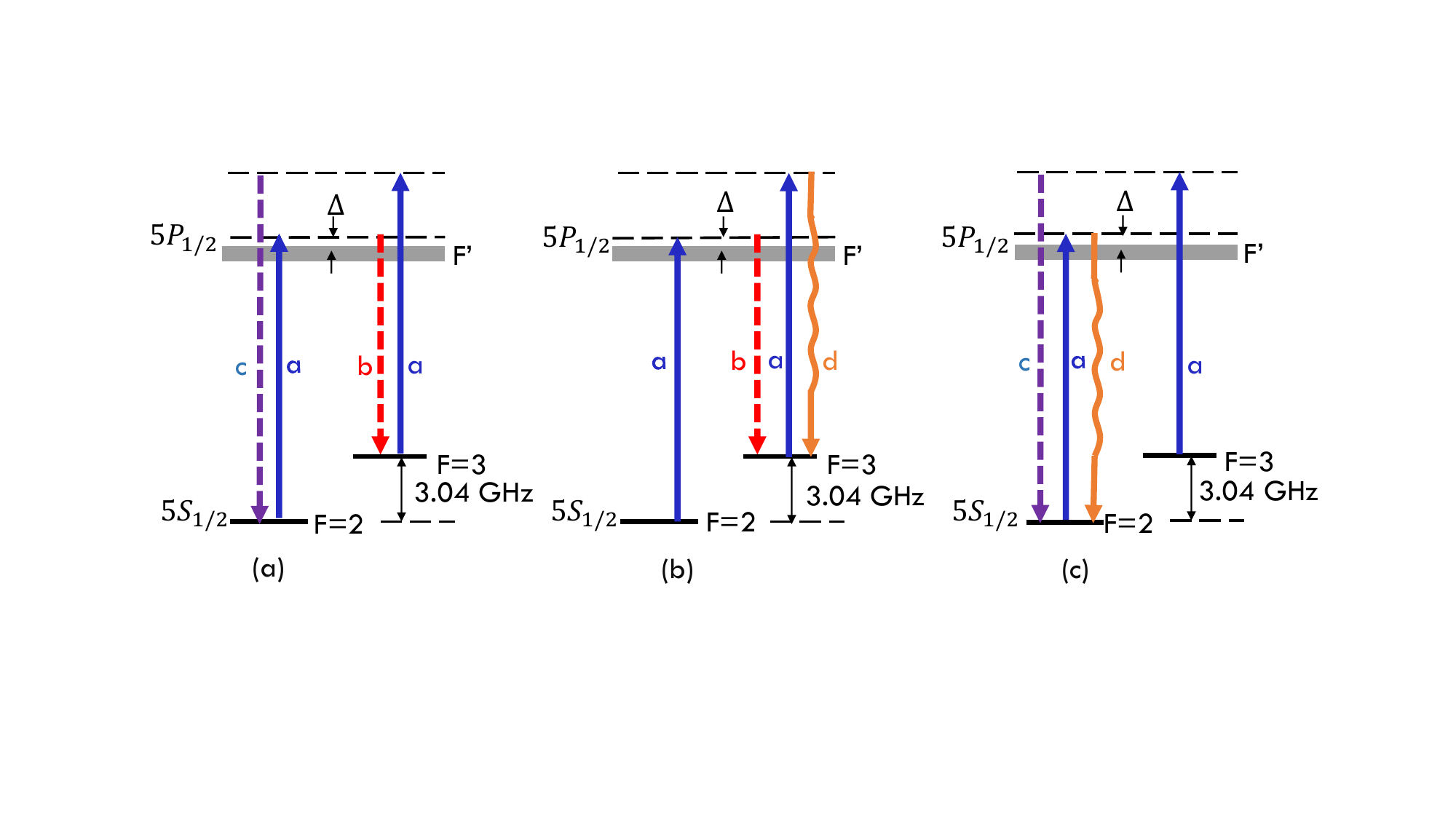}
    \caption{Energy-level diagrams of the dominant and parasitic FWM processes. Fig.~\ref{fig:energy_levels}(a) shows the dominant $\chi^{(3)}$ SFWM process in a double-$\Lambda$ configuration, where two pump photons (field $a$) drive transitions from both $F=2$ and $F=3$ ground states to $F'$ and generate correlated probe ($b$) and conjugate ($c$) fields. Figs.~\ref{fig:energy_levels}(b) and (c) show parasitic Raman--spontaneous emission channels: in Fig.~\ref{fig:energy_levels}(b), a pump-driven Raman process from $F=2$ produces $b$ and transfers population to $F=3$, followed by pump-induced excitation and spontaneous emission generating a parasitic photon ($d$) that replaces a true conjugate; in Fig.~\ref{fig:energy_levels}(c), atoms in $F=3$ undergo a pump-driven Raman transition producing $c$, followed by spontaneous decay generating a parasitic photon ($d$). Residual $F=3$ population due to lack of optical pumping enables these pathways alongside SFWM, leading to false-partner events in the detected coincidences arising from non-phase-matched emission into the probe and conjugate detection modes.
    }
    \label{fig:energy_levels}    
\end{figure}

The relevant field definitions and level structure are summarized in Fig.~\ref{fig:energy_levels}, where $a$ denotes the pump field, $b$ the probe (signal), $c$ the conjugate (idler), and $d$ parasitic fields generated via spontaneous emission processes. The energy level diagram of the $^{85}$Rb D1 line consists of the $5S_{1/2}$ ground manifold, split into $F=2$ and $F=3$ hyperfine states separated by 3.04~GHz, and the $5P_{1/2}$ excited manifold $F'$. The pump is detuned by $\Delta = 800$ MHz from the $5S_{1/2} \rightarrow 5P_{1/2}$ resonance.

The dominant process is SFWM arising from a $\chi^{(3)}$ nonlinearity in a double-$\Lambda$ configuration shown in Fig.~\ref{fig:energy_levels}(a), in which two pump photons generate time--frequency--correlated probe ($b$) and conjugate ($c$) photon pairs under the phase-matching condition $\Delta k = 2k_a - k_b - k_c = 0$. In the Doppler-broadened $^{85}$Rb vapor, the $\chi^{(3)}$ gain acquires an inhomogeneous width $\Gamma_D \sim k_a \bar{v}$, where $\bar{v} = \sqrt{2k_B T / m}$ is the most probable atomic speed. As a result, the SFWM gain spectrum extends over the GHz range, greatly exceeding the homogeneous linewidth $\Gamma_0$. This Doppler-broadened $\chi^{(3)}$ response relaxes the phase-matching constraint, providing a finite tolerance window around $\Delta k = 0$. In the absence of a dedicated repump field, incomplete optical pumping leaves a residual population in the $F=3$ ground state, giving rise to two parasitic FWM channels shown in Figs.~\ref{fig:energy_levels}(b) and (c). These parasitic processes do not satisfy the phase-matching condition and therefore produce uncorrelated photons~\cite{Du2008S}.

In the first parasitic FWM process shown in Fig.~\ref{fig:energy_levels}(b), a pump photon drives a stimulated Raman transition from the $F=2$ ground state through the excited state $F'$, generating a probe photon ($b$) and leaving the atom in the $F=3$ ground state. Due to the lack of optical pumping, atoms in the $F=3$ ground state can be re-excited by the pump to the excited state, followed by spontaneous decay back to $F=3$, emitting a parasitic photon ($d$) that is detected in place of a true conjugate. Because the Raman step generating $b$ is not accompanied by a phase-matched conjugate-generation process, no correlated partner photon is produced, and the resulting coincidences correspond to false-partner events at the single-photon level. 

In the second parasitic FWM process shown in Fig.~\ref{fig:energy_levels}(c), atoms initially residing in the $F=3$ ground state undergo a Raman transition that generates a conjugate photon ($c$) via stimulated emission into the $F=2$ ground state. A subsequent pump interaction excites the atom to the excited state, followed by spontaneous decay back to $F=3$, producing a parasitic photon ($d$) that is registered in the trigger channel as a false probe.

Coincidence analysis is performed in a start--stop configuration, with the probe ($b$) in channel 2 defining the trigger and channel 1 recording the partner photon. For the dominant SFWM process, channel 1 detects the conjugate ($c$), producing a coincidence peak at $\tau = 0$. In contrast, the two parasitic FWM processes give rise to additional features on both sides of the time-delay plot due to the sequential and probabilistic generation of $c$ and $d$. Because channel 2 serves as the trigger, events involving the parasitic photon ($d$) can lead to misidentified partner detections. As a result, coincidences appear at both $\tau > 0$ and $\tau <0$. This results in approximately symmetric contributions about $\tau = 0$ in the measured arrival-time delay distribution. \\

The nonlinear atomic medium is driven by a classical pump field at frequency $\omega_{\mathrm{p}}$ and consists of two hyperfine ground states $|1\rangle \equiv |5S_{1/2}, F=2\rangle$ and $|2\rangle \equiv |5S_{1/2}, F=3\rangle$. The biphoton temporal wavefunction generated via SFWM is
\begin{equation}
\Psi(t_{\mathrm{pr}},t_{\mathrm{c}})
=
\psi(\tau)\,
e^{-i(\omega_{\mathrm{pr}} t_{\mathrm{pr}} + \omega_{\mathrm{c}} t_{\mathrm{c}})},
\end{equation}
where $\tau = t_{\mathrm{c}} - t_{\mathrm{pr}}$ is the relative delay between the probe and conjugate photons, and $\omega_{\mathrm{pr}}$ and $\omega_{\mathrm{c}}$ are the probe and conjugate angular frequencies, respectively, satisfying energy conservation
\begin{equation}
2\omega_{\mathrm{p}} = \omega_{\mathrm{pr}} + \omega_{\mathrm{c}}.
\end{equation}
The relative temporal wavefunction is
\begin{equation}
\psi(\tau)
=
\frac{1}{2\pi}
\int d\omega\, \kappa(\omega)\,\Phi(\omega)\,e^{-i\omega\tau},
\end{equation}
where $\kappa(\omega)$ is the nonlinear coupling coefficient and
\begin{equation}
\Phi(\omega)
=
\mathrm{sinc}\!\left(\frac{\Delta k L}{2}\right)
\ e^{\! i(k_{pr} + k_{c}) L/{2}},
\end{equation}
is the longitudinal phase-matching function, with phase mismatch
$\Delta k = 2k_{\mathrm{p}} - k_{\mathrm{pr}} - k_{\mathrm{c}}$ and $L$ the
interaction length. The wave numbers of the probe and conjugate fields, as well as the linear and third-order nonlinear susceptibilities, are given by
\begin{equation}
k_{\mathrm{pr}}(\omega_{\mathrm{pr}})
=
\frac{\omega_{\mathrm{pr}}}{c}\!\left(1 + \frac{\chi_{\mathrm{pr}}(\omega_{\mathrm{pr}})}{2}\right),
\quad
k_{\mathrm{c}}(\omega_{\mathrm{c}})
=
\frac{\omega_{\mathrm{c}}}{c}\!\left(1 + \frac{\chi_{\mathrm{c}}(\omega_{\mathrm{c}})}{2}\right).
\end{equation}
\begin{equation}
\chi_{\mathrm{pr}}(\omega)
=
\int dv\, f(v)\,
\frac{N}{\epsilon_0 \hbar}
\sum_{F'=2,3}
\frac{|\mu_{2F'}|^2}{\Delta_{2F'}(\omega,v) - i\gamma_{F'}},
\end{equation}
\begin{equation}
\chi_{\mathrm{c}}(\omega)
=
\int dv\, f(v)\,
\frac{N}{\epsilon_0 \hbar}
\sum_{F'=2,3}
\frac{|\mu_{1F'}|^2}{\Delta_{1F'}(\omega,v) - i\gamma_{F'}},
\end{equation}
\begin{equation}
\chi^{(3)}(\omega)
=
\int dv\, f(v)\,
\frac{N}{\epsilon_0 \hbar^3}
\sum_{F'=2,3}
\frac{
|\mu_{2F'}|^2\,|\mu_{1F'}|^2
}{
\big(\Delta_{2F'}(\omega,v) - i\gamma_{F'}\big)
\big(\Delta_{32}(\omega,v) - i\gamma_{32}\big)
\big(\Delta_{1F'}(\omega,v) - i\gamma_{F'}\big)
}.
\end{equation}

Under the present experimental conditions, the excited-state hyperfine structure is not spectrally resolved due to Doppler broadening and the one-photon detuning ($\sim 800~\mathrm{MHz}$). We therefore approximate the excited manifold by an effective state $|e\rangle$. The susceptibilities then reduce to
\begin{equation}
\chi_{\mathrm{pr}}(\omega)
=
\int dv\, f(v)\,
\frac{N}{\epsilon_0 \hbar}
\frac{|\mu_{2e}|^2}{\Delta_{\mathrm{pr}}(\omega,v) - i\gamma},
\end{equation}
\begin{equation}
\chi_{\mathrm{c}}(\omega)
=
\int dv\, f(v)\,
\frac{N |\Omega_{\mathrm{p}}|^2}{4\epsilon_0 \hbar \Delta^2}
\frac{|\mu_{1e}|^2}{\Delta_{\mathrm{c}}(\omega,v) - i\gamma},
\end{equation}
\begin{equation}
\chi^{(3)}(\omega)
=
\int dv\, f(v)\,
\frac{N |\Omega_{\mathrm{p}}|^2}{4\epsilon_0 \hbar^3 \Delta^2}
\frac{
|\mu_{2e}|^2\,|\mu_{1e}|^2
}{
\big(\Delta_{\mathrm{c}}(\omega,v) - i\gamma\big)
\big(\Delta_{32}(\omega,v) - i\gamma_{32}\big)
\big(\Delta_{\mathrm{pr}}(\omega,v) - i\gamma\big)
},
\end{equation}
where the 1D Maxwell-Boltzmann distribution is
\begin{equation}
f(v)
=
\sqrt{\frac{m}{2\pi k_B T}}
\exp\!\left(-\frac{mv^2}{2k_B T}\right),
\end{equation}
and $\Delta = \omega_{\mathrm{p}} - \omega_e$ is the one-photon pump detuning with $k_B$ the Boltzmann constant, $T$ the cell temperature, $N$ the atomic density, and the Rabi frequency of the pump is $\Omega_{\mathrm{p}} = \mu_0 E_{\mathrm{p}}/\hbar$. The velocity-dependent detunings are
\begin{equation}
\Delta_{\mathrm{pr}}(\omega,v)
=
\omega_{\mathrm{pr}} - \omega_{e} - k_{\mathrm{pr}} v,
\end{equation}
\begin{equation}
\Delta_{\mathrm{c}}(\omega,v)
=
\omega_{\mathrm{c}} - \omega_{e} - k_{\mathrm{c}} v,
\end{equation}
\begin{equation}
\Delta_{32}(\omega,v)
=
\omega_{\mathrm{pr}} - \omega_{\mathrm{c}} - \omega_{32}
- (k_{\mathrm{pr}} - k_{\mathrm{c}}) v.
\end{equation}
The dipole matrix elements are $|\mu_{ie}|^2 = |\mu_0|^2 C_{ie}$, where $C_{ie}$ accounts for excited-state hyperfine structure and reduces to $C_{ie}=1$ ($i=1,2$) in the unresolved-$F'$ limit. The reduced matrix element is related to the natural linewidth by
\begin{equation}
|\mu_0|^2 = \frac{3\pi\epsilon_0\hbar c^3}{\omega_0^3}\,\Gamma,
\end{equation}
where $\Gamma = 2\pi \times 5.75~\mathrm{MHz}$ is the natural linewidth of
the $^{85}$Rb D1 transition.

The nonlinear coupling coefficient is
\begin{equation}
\kappa(\omega) \propto \chi^{(3)}(\omega)\,E_{\mathrm{p}}^{2},
\end{equation}
and the biphoton coincidence counts are calculated from%
\begin{equation}
C(\tau)
=
\beta\,|\psi(\tau)|^2\,\Delta t\,T,
\end{equation}
where $\beta$ is the detection efficiency, $\Delta t$ is the bin width, and $T$ is the acquisition time.\\

\begin{table}[]
\caption{Experimental and derived parameters used in the SFWM model in warm $^{85}$Rb vapor. Frequencies are expressed in angular units where applicable.\\}
\label{tab:parameters}
\centering
\begin{tabular}{lll}
\hline\hline
Parameter & Description & Value \\
\hline

$T$ & Cell temperature & $92~^\circ\mathrm{C}$ \\

$L$ & Interaction length & $12.5~\mathrm{mm}$ \\

$\lambda$ & Optical wavelength (D1 line) & $795~\mathrm{nm}$ \\

$\Gamma$ & Natural linewidth & $2\pi \times 5.75~\mathrm{MHz}$ \\

$\Delta$ & Pump detuning & $2\pi \times 800~\mathrm{MHz}$ \\

$\Delta_{32}$ & Two-photon detuning & $2\pi \times 4~\mathrm{MHz}$ \\

$P_{\mathrm{p}}$ & Pump power & $5~\mathrm{mW}$ \\

$d_{\mathrm{p}},\, d_{\mathrm{pr}}$ & Beam diameters (pump/probe) & $1 / 0.5~\mathrm{mm}$ \\

$\Delta t_{\mathrm{bin}}$ & Time-bin width & $250~\mathrm{ps}$ \\

$\tau_{\mathrm{win}}$ & Acquisition window & $\pm 30~\mathrm{ns}$ \\

$\gamma$ & Excited-state decay rate & $\Gamma/2$ \\

$\mu_{0}$   & Reduced dipole matrix element (D1 line) & $2.54 \times 10^{-29}\,\mathrm{C\cdot m}$ \\



$\gamma_{32}$ & Ground-state decoherence rate & $2\pi \times 100~\mathrm{kHz}$ \\

$N$ & Atomic number density & $6 \times 10^{18}~\mathrm{m^{-3}}$ \\

$\Omega_{\mathrm{p}}$ & Pump Rabi frequency & $ 2\pi \times 100~\mathrm{MHz}$ \\

\hline\hline
\end{tabular}
\end{table}
In addition to the SFWM emission, two parasitic FWM contributions arise from two pump-induced spontaneous Raman processes connecting the hyperfine ground states, as illustrated in Figs.~\ref{fig:energy_levels}(b) and (c). Within this effective description, the Doppler-averaged linear and nonlinear susceptibilities describe contributions that populate the detection channels without generating phase-matched correlated photon pairs. 
\begin{figure}[h]
\centering
\includegraphics[width=0.48\textwidth]{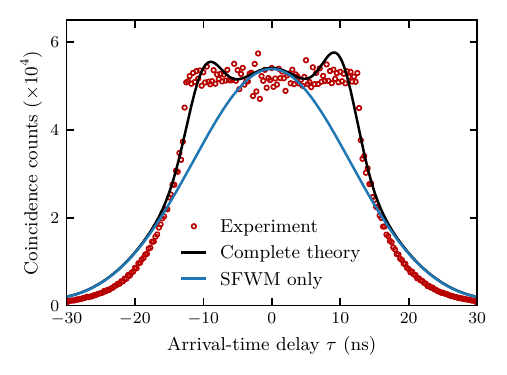}
\caption{
Coincidence counts as a function of the arrival-time delay between the probe and conjugate photons (red open circles), shown together with the theoretical prediction including only the dominant SFWM process (blue curve) and the complete theoretical model (black curve).
}
\label{fig:SFWM}
\end{figure}
These channels are described by Doppler-averaged linear susceptibilities given by
\begin{equation}
\chi^{(1)}_j(\omega)
=
\int dv\, f(v)\,
\mathcal{F}_{\chi^{(1)}_j}(\omega,v)
\label{eq:S10}
\end{equation}
where $j \in \{F=2, F=3\}$ labels the two hyperfine Raman pathways.

The corresponding spectral amplitudes are
\begin{equation}
S_j(\omega)
\propto
\chi^{(1)}_j(\omega)\,E_{\mathrm{p}},
\label{eq:S11}
\end{equation}
where $E_{\mathrm{p}}$ is the pump field amplitude. To account for the experimentally observed temporal offsets, a 
frequency-domain phase factor is introduced,
\begin{equation}
S_j(\omega)
\rightarrow
S_j(\omega)\,e^{-i\omega\tau_j},
\label{eq:S12}
\end{equation}
which shifts the temporal response by $\tau_j$. The corresponding temporal amplitudes are obtained via Fourier transform as
\begin{equation}
\psi_j(\tau)
=
\frac{1}{2\pi}
\int d\omega\, S_j(\omega)\,e^{-i\omega\tau},
\end{equation}
and the associated probability densities are
\begin{equation}
p_j(\tau)
=
|\psi_j(\tau)|^2 .
\label{eq:S13}
\end{equation}

The full model consists of the SFWM contribution together with the two parasitic Raman-assisted scattering channels, each involving Raman scattering followed by spontaneous emission,
\begin{equation}
C(\tau)
=
\Delta t_{\mathrm{bin}}\,T
\left[
\beta\,|\psi(\tau)|^2
+
\sum_{j\in\{F=2,F=3\}}
\beta_j\,p_j(\tau)
\right],
\label{eq:S14}
\end{equation}
where $\Delta t_{\mathrm{bin}}$ is the time-bin width and $T$ is the acquisition time. The coefficient $\beta_{\mathrm{FWM}}$ is fixed by the central coincidence peak, while the coefficients $\beta_j$ are determined from the amplitudes of the two Raman side peaks. The amplitudes $\beta_{F=2}$ and $\beta_{F=3}$ are extracted by matching the experimentally observed side peaks at $\tau_{F=2}=+10.5~\mathrm{ns}$ and $\tau_{F=3}=-10.5~\mathrm{ns}$. Any asymmetry in the measured $C(\tau)$ about $\tau=0$ directly reflects the ratio $\beta_{F=2}/\beta_{F=3}$, providing a sensitive probe of the residual ground-state population imbalance and optical pumping efficiency. As shown in Fig.~\ref{fig:SFWM}, the SFWM-only model predicts a smooth coincidence profile, whereas inclusion of the two Raman-assisted scattering channels reproduces the shoulders near $\tau=\pm10.5~\mathrm{ns}$ observed experimentally, demonstrating that these additional processes are responsible for the modified temporal correlation profile. The experimental parameters used in the theoretical model are summarized in Table~\ref{tab:parameters}, with some values taken from Ref.~\cite{SteckRb85S}.

\section*{Supplemental Note B: Theoretical model of Sub-Poissonian Photon Statistics in Individual Fields}

The purpose of this Supplemental Note is to develop an effective theoretical framework for the observed sub-Poissonian photon statistics of the individual probe and conjugate fields generated by SFWM in a warm rubidium vapor. Rather than deriving a complete microscopic theory of warm-vapor SFWM~\cite{Glorieux2010}, we identify the essential physical ingredients responsible for the observed negative Mandel-$Q$ parameter.

The model is motivated by several experimentally established features of the present system. SFWM in warm atomic vapors generates multiple probe--conjugate spatio-temporal Schmidt-mode pairs that share a \textit{finite pump-dressed atomic coherence and therefore compete for the available nonlinear gain through self- and cross-saturation}. The polarization-maintaining (PM) fibers project the generated light onto a particular spatial superposition of these modes, while the finite counting-bin duration integrates the detected photon flux over a finite temporal interval.

Motivated by these considerations, we introduce a \textit{phenomenological} occupation-dependent nonlinear extension of the conventional undepleted-pump SFWM model. The operator-valued saturation law is inspired by the well-established theory of multimode gain saturation and mode competition in optical amplifiers, particularly traveling-wave amplifiers~\cite{Napartovich2007}, which provide the closest analogue to the present warm-vapor SFWM system. Promoting the classical modal intensities to quantum number operators introduces the essential non-Gaussian element of the model, leading to the suppression of large photon-number fluctuations and enabling sub-Poissonian photon statistics.

The derivation proceeds by first reviewing the conventional multimode SFWM Hamiltonian and its thermal marginal statistics. We then introduce the occupation-dependent nonlinear Hamiltonian, discuss collective mode competition and PM-fiber projection, and finally derive an effective birth--death model that predicts a stationary pair-number distribution with a negative Mandel-$Q$ parameter.

\subsection*{B1. Conventional multimode SFWM theory and justification for a nonlinear extension}

We begin with the conventional undepleted-pump description of spontaneous four-wave mixing (SFWM). Let $\mu$ denote the paired spatio-temporal Schmidt modes of the probe and conjugate fields, with annihilation operators $\hat a_\mu$ and $\hat b_\mu$ satisfying
\[
[\hat a_\mu,\hat a_\nu^\dagger]=\delta_{\mu\nu},
\qquad
[\hat b_\mu,\hat b_\nu^\dagger]=\delta_{\mu\nu}.
\]
Within the undepleted-pump approximation, the interaction Hamiltonian is
\[
\hat H_{\rm lin}
=
i\hbar\sum_\mu
\kappa_{\mu0}
\left(
\hat a_\mu^\dagger\hat b_\mu^\dagger
-
\hat a_\mu\hat b_\mu
\right),
\]
which generates independent two-mode squeezed-vacuum states in the Schmidt basis. Consequently, after tracing over one member of each photon pair, every individual Schmidt mode exhibits thermal photon-number statistics with a non-negative Mandel-$Q$ parameter. This constitutes the standard theoretical description of SFWM.

The thermal-marginal prediction relies on three assumptions: (i) the pump remains undepleted, (ii) the atomic medium is unchanged during the interaction, and (iii) the pair-generation amplitudes $\kappa_{\mu0}$ are independent of the occupation of the generated modes. Under the present experimental conditions, the first two assumptions remain well justified. The third assumption, however, may no longer hold in a finite warm-vapor medium. Since multiple SFWM spatio-temporal modes share the same pump-dressed atomic coherence, they compete for a common nonlinear gain resource. As the occupation of one mode increases, the gain available to overlapping modes is expected to decrease, providing the physical motivation for introducing an occupation-dependent nonlinear gain.

A widely used semiclassical description of multimode gain saturation is
\[
\kappa_\mu(\{I_\nu\})
=
\frac{\kappa_{\mu0}}
{1+\dfrac{1}{I_{\rm sat}}
\sum_\nu
C_{\mu\nu}I_\nu},
\]
where $I_\nu$ is the intensity of mode $\nu$, $C_{\mu\nu}\ge0$ quantifies the strength of gain competition between modes, and $I_{\rm sat}$ is the saturation intensity. This expression follows directly from the standard gain-saturation model
\[
g=
\frac{g_0}{1+I/I_{\rm sat}}.
\]
For weak saturation,
\[
\frac{1}{1+x}=1-x+O(x^2),
\]
which gives
\[
\kappa_\mu(\{I_\nu\})
\simeq
\kappa_{\mu0}
\left[
1-
\frac{1}{I_{\rm sat}}
\sum_\nu
C_{\mu\nu}I_\nu
\right].
\]
This linearized expression provides the classical motivation for the quantum occupation-dependent model introduced in the following section.

\subsection*{B2. Occupation-dependent nonlinear Hamiltonian}

To incorporate occupation-dependent gain into the quantum description, the classical modal intensities are promoted to quantum number operators. We define the pair-occupation operator as
\[
\hat{N}_\nu=
\frac{\hat{a}_\nu^\dagger\hat{a}_\nu+\hat{b}_\nu^\dagger\hat{b}_\nu}{2},
\]
which has eigenvalue $n_\nu$ within the ideal twin-number subspace
$\ket{n_\nu,n_\nu}$. The effective pair-generation coefficient is then written as
\[
\hat{\kappa}_\mu=
\kappa_{\mu0}
\left[
1-
\frac{1}{N_{\rm sat}}
\sum_\nu
C_{\mu\nu}\hat{N}_\nu
\right],
\]
where $N_{\rm sat}$ denotes the effective saturation occupation and
$\mathbf{C}=(C_{\mu\nu})$ is the mode-competition matrix. The matrix elements are defined by
\[
C_{\mu\nu}
=
\int
u_\mu^*(\mathbf r,t)\,
W(\mathbf r,t)\,
u_\nu(\mathbf r,t)\,
d^3r\,dt,
\]
where $u_\mu(\mathbf r,t)$ is the $\mu$th spatio-temporal Schmidt mode and $W(\mathbf r,t)$ represents the spatial and temporal distribution of the pump-dressed atomic coherence, or more generally the nonlinear susceptibility responsible for the SFWM process. Physically, two Schmidt modes compete only to the extent that they overlap within the same nonlinear gain region. \textit{Consequently, $C_{\mu\nu}$ quantifies the degree to which the occupation of Schmidt mode $\nu$ reduces the pump-dressed atomic coherence available for pair generation into Schmidt mode $\mu$}. The diagonal elements of $\mathbf C$ describe self-saturation, whereas the off-diagonal elements characterize cross-saturation between different Schmidt modes.

Since the occupation operator $\hat N_\nu$ does not commute with the pair-creation and pair-annihilation operators, the operator ordering must be specified explicitly. A manifestly Hermitian form of the nonlinear interaction Hamiltonian is
\[
\hat H_{\rm nl}
=
i\hbar\sum_\mu
\kappa_{\mu0}
\left[
\hat a_\mu^\dagger
\hat b_\mu^\dagger
f_\mu(\hat{\mathbf N})
-
f_\mu(\hat{\mathbf N})
\hat a_\mu
\hat b_\mu
\right],
\]
where
\[
f_\mu(\hat{\mathbf N})
=
1-
\frac{1}{N_{\rm sat}}
\sum_\nu
C_{\mu\nu}\hat N_\nu .
\]
In this form, the available nonlinear gain is evaluated before the creation or annihilation of each photon pair, thereby ensuring Hermiticity of the interaction Hamiltonian. Since the gain function $f_\mu(\hat{\mathbf N})$ explicitly depends on the photon-number operators, the Hamiltonian is no longer quadratic in the field operators. Consequently, the generated quantum state is non-Gaussian, in contrast to the conventional two-mode squeezed-vacuum state predicted by the linear SFWM Hamiltonian.

The occupation-dependent gain model introduced here is a \textit{phenomenological} quantum extension of the classical multimode gain-saturation model. Although it is not derived from a microscopic theory of warm-vapor SFWM, it captures the essential physics of mode competition arising from a \textit{shared finite nonlinear gain resource}. As shown below, this nonlinear extension naturally provides a physically motivated mechanism for the experimentally observed sub-Poissonian photon statistics of the individual probe and conjugate fields.

\subsection*{B3. Effective single-mode description}

To gain physical insight into the nonlinear interaction, we first consider an effective probe--conjugate mode pair described by the operators $\hat A$ and $\hat B$. In this limit, the occupation-dependent gain function introduced above reduces to
\[
f(\hat N)=1-\frac{\hat N}{N_{\rm sat}},
\]
and the nonlinear interaction Hamiltonian takes the form
\[
\hat H_{\rm eff}
=
i\hbar\kappa_0
\left[
\hat A^\dagger
\hat B^\dagger
f(\hat N)
-
f(\hat N)
\hat A
\hat B
\right],
\]
where
\[
\hat N
=
\hat A^\dagger\hat A
=
\hat B^\dagger\hat B
\]
is the photon-number operator of the twin-mode pair. Within the twin-number basis,
\[
\hat N\ket{n,n}=n\ket{n,n},
\]
the pair-creation operator acts as
\[
\hat A^\dagger
\hat B^\dagger
f(\hat N)
\ket{n,n}
=
(n+1)
\left(
1-\frac{n}{N_{\rm sat}}
\right)
\ket{n+1,n+1}.
\]
The transition amplitude therefore contains two competing factors,
\[
(n+1)
\qquad\text{and}\qquad
\left(1-\frac{n}{N_{\rm sat}}\right),
\]
which represent bosonic enhancement and occupation-dependent gain saturation, respectively. In the conventional undepleted-pump SFWM Hamiltonian, the pair-generation dynamics are governed solely by the bosonic enhancement factor $(n+1)$, leading to the familiar geometric photon-number distribution of the two-mode squeezed vacuum. In the present model, however, the occupation-dependent saturation factor progressively reduces the effective nonlinear gain as the modal occupation increases. Consequently, the transition probabilities between successive twin-number states are no longer determined solely by bosonic enhancement but are modified by the competition between stimulated pair generation and gain saturation. These modified transition probabilities form the basis of the effective birth--death model developed in the following section, from which the stationary photon-number distribution and the corresponding Mandel-$Q$ parameter are obtained.

\subsection*{B4. Collective competition eigenmodes and PM-fiber projection}

The Schmidt basis diagonalizes the conventional linear SFWM Hamiltonian introduced in Sec.~B1, so that the individual Schmidt modes evolve independently. The introduction of the occupation-dependent gain, however, couples different Schmidt modes through the competition matrix $\mathbf C$, and the Schmidt basis is therefore no longer the natural basis of the nonlinear interaction.

Since the competition matrix $\mathbf C$ is Hermitian (and, for the present model, real and symmetric), there exists a unitary transformation $\mathbf U$ such that
\[
\mathbf U^\dagger
\mathbf C
\mathbf U
=
\mathrm{diag}
(\lambda_1,\lambda_2,\ldots).
\]
The eigenvectors of $\mathbf C$ define a new set of orthogonal collective modes,
\[
\hat A_i
=
\sum_\mu
U_{\mu i}\hat a_\mu,
\qquad
\hat B_i
=
\sum_\mu
U_{\mu i}\hat b_\mu.
\]
These collective competition eigenmodes diagonalize the nonlinear gain competition rather than the linear SFWM interaction. Physically, each collective mode represents an independent channel competing for the finite pump-dressed atomic coherence, with the strength of the nonlinear gain saturation determined by the corresponding eigenvalue $\lambda_i$.

In the experiment, the probe and conjugate fields are collected by polarization-maintaining (PM) fibers. In general, the guided fiber mode does not coincide with an individual Schmidt mode of the generated SFWM field. Instead, the detected fields correspond to coherent superpositions of Schmidt modes,
\[
\hat A_{\rm det}
=
\sum_\mu c_\mu \hat a_\mu,
\qquad
\hat B_{\rm det}
=
\sum_\mu d_\mu \hat b_\mu,
\]
with normalization
\[
\sum_\mu |c_\mu|^2=1,
\qquad
\sum_\mu |d_\mu|^2=1.
\]
Expressed in the collective competition basis,
\[
\hat A_{\rm det}
=
\sum_i e_i \hat A_i,
\qquad
\hat B_{\rm det}
=
\sum_i f_i \hat B_i,
\]
where
\[
c_\mu
=
\sum_i e_i U_{\mu i},
\qquad
d_\mu
=
\sum_i f_i U_{\mu i},
\]
and $U_{\mu i}$ is the unitary transformation that diagonalizes the competition matrix.

When the PM-fiber projection is dominated by a single collective competition eigenmode,
\[
|e_1|^2 \gg |e_{i>1}|^2,
\qquad
|f_1|^2 \gg |f_{i>1}|^2,
\]
the detected fields reduce to
\[
\hat A_{\rm det}
\simeq
\hat A_1
=
\sum_\mu U_{\mu1}\hat a_\mu,
\qquad
\hat B_{\rm det}
\simeq
\hat B_1
=
\sum_\mu U_{\mu1}\hat b_\mu.
\]
Under this condition, the multimode nonlinear Hamiltonian reduces to
\[
\hat H_{\rm eff}
=
i\hbar\kappa_0
\left[
\hat A_1^\dagger
\hat B_1^\dagger
f(\hat N_1)
-
f(\hat N_1)
\hat A_1
\hat B_1
\right],
\]
where
\[
f(\hat N_1)
=
1-
\frac{\hat N_1}{N_{\rm sat}}.
\]
Thus, when the PM-fiber projection is dominated by a single collective competition eigenmode, the effective single-mode theory developed in Sec.~B3 provides an accurate description of the experimentally detected fields.

\subsection*{B5. Coarse-grained photon counting}

The time-resolved photon-counting measurements performed in the experiment do not detect instantaneous photon numbers but rather the photon flux integrated over a finite counting interval of duration $T$. The measured photon-number operators are therefore
\[
\hat N_A^{(k)}
=
\int_{t_k}^{t_k+T}
\hat A^\dagger(t)\hat A(t)\,dt,
\qquad
\hat N_B^{(k)}
=
\int_{t_k}^{t_k+T}
\hat B^\dagger(t)\hat B(t)\,dt.
\]
A finite counting window corresponds to an integrated photon-number measurement rather than a projection onto a single temporal Schmidt mode. Consequently, the effective single-temporal-mode approximation is justified only when one temporal coherence mode dominates the selected integration interval or when the detector bandwidth is sufficiently narrow that the detected field is effectively single mode. Otherwise, several temporal coherence modes contribute simultaneously to the measured photon statistics, and the complete multimode description developed above must be retained.

\subsection*{B6. Diagonal pair-number rate model and stationary photon-number statistics}

To obtain an analytically tractable description of the nonlinear population dynamics, we consider the regime in which atomic dephasing is sufficiently strong that coherences between different twin-number states decay much faster than the population dynamics. Under this condition, the density operator is effectively diagonal in the twin-number basis, and the nonlinear quantum dynamics reduce to an effective birth--death process for the pair number $n$.

The corresponding birth and death processes are
\[
n
\xrightarrow{\,R_{+}(n)\,}
n+1,
\qquad
n
\xrightarrow{\,R_{-}(n)\,}
n-1,
\]
with transition rates
\[
R_{+}(n)
=
\Gamma_{+}(n+1)
\left(
1-\frac{n}{N_{\rm sat}}
\right),
\qquad
R_{-}(n)
=
\Gamma_{-}n,
\]
where $\Gamma_{+}$ is the unsaturated pair-generation rate and $\Gamma_{-}$ is the effective loss (or escape) rate. The factor $(n+1)$ represents the familiar bosonic enhancement associated with stimulated pair generation, whereas the factor $1-n/N_{\rm sat}$ describes the reduction of the available nonlinear gain arising from occupation-dependent saturation. The linear loss rate reflects independent removal of photons from the populated mode.

The birth--death model is an effective description of the nonlinear Hamiltonian introduced in Sec.~B2. Although it is not an exact reduction of the full quantum dynamics, it retains the essential physical ingredients of bosonic enhancement, occupation-dependent gain saturation, and linear loss while permitting an analytical determination of the stationary photon-number distribution.

In the stationary state, the birth and death processes satisfy the detailed-balance condition
\[
P_{n+1}R_{-}(n+1)
=
P_nR_{+}(n),
\]
where $P_n$ denotes the stationary probability of finding $n$ photon pairs. Substituting the transition rates gives
\[
\frac{P_{n+1}}{P_n}
=
r
\left(
1-\frac{n}{N_{\rm sat}}
\right),
\qquad
r\equiv
\frac{\Gamma_{+}}{\Gamma_{-}}.
\]
Iterating this recurrence relation yields
\[
P_n
=
P_0
\,r^n
\prod_{k=0}^{n-1}
\left(
1-\frac{k}{N_{\rm sat}}
\right),
\qquad
0\le n\le N_{\rm sat},
\]
where the normalization constant $P_0$ satisfies
\[
\sum_{n=0}^{N_{\rm sat}}P_n=1.
\]
In the conventional undepleted-pump SFWM model, the ratio
\[
P_{n+1}/P_n=r
\]
is independent of $n$, giving the familiar geometric photon-number distribution of the two-mode squeezed vacuum. In contrast, occupation-dependent saturation causes the ratio $P_{n+1}/P_n$ to decrease monotonically with increasing occupation number, progressively suppressing the high-photon-number tail of the distribution relative to the thermal case.

Analytical insight can be obtained in the limit of large saturation occupation,
$N_{\rm sat}\gg1$. Expanding $\ln P_n$ about its maximum and retaining terms up to second order yields an approximately Gaussian distribution with variance
\[
\operatorname{Var}(n)
\simeq
\frac{N_{\rm sat}}{r},
\]
and mean photon-pair number
\[
\langle n\rangle
\simeq
N_{\rm sat}
\left(
1-\frac{1}{r}
\right)
=
N_{\rm sat}
\frac{r-1}{r}.
\]
The corresponding Mandel-$Q$ parameter is
\[
Q
=
\frac{\operatorname{Var}(n)-\langle n\rangle}
{\langle n\rangle}
\simeq
\frac{2-r}{r-1}.
\]
The model therefore predicts
\[
Q<0,
\qquad
\text{for}
\qquad
r>2.
\]
Thus, within the present effective model, the photon-number fluctuations become sub-Poissonian once the unsaturated pair-generation rate exceeds approximately twice the effective loss rate ($r>2$). As the gain ratio increases further, the stationary photon-number distribution becomes progressively more number stabilized. Some representative values are
\[
r=2:\quad Q\simeq0,\qquad
r=3:\quad Q\simeq-\frac12,\qquad
r=5:\quad Q\simeq-\frac34,
\]
and
\[
r\rightarrow\infty:\qquad
Q\rightarrow-1,
\]
corresponding to an increasingly narrow photon-number distribution centered near the saturation occupation. Notably, the generation-to-removal ratio of $r=5$ gives rise to $Q\simeq-3/4$, consistent with the magnitude of the sub-Poissonian statistics observed experimentally. The relatively low pump power used in our experiment ($\sim 5~\mathrm{mW}$) may further favor operation outside the ideal undepleted-pump approximation, making finite-resource saturation and pump-mediated nonlinear feedback plausible under our experimental conditions.

\subsection*{B7. Discussion}

The effective theoretical framework developed above provides a physical mechanism for the emergence of sub-Poissonian photon-number statistics in warm-vapor SFWM. In the multimode case, the occupation-dependent pair-generation rate is described by
\[
R_{\mu,+}(\mathbf n)
=
\Gamma_{\mu,+}(n_\mu+1)
\left[
1-
\frac{1}{N_{\rm sat}}
\sum_\nu C_{\mu\nu}n_\nu
\right],
\]
where the competition matrix $C_{\mu\nu}$ quantifies the extent to which different Schmidt modes share the same pump-dressed atomic gain resource. Consequently, pair generation in one mode reduces the available nonlinear gain for competing modes whenever $C_{\mu\nu}>0$, giving rise to nonlinear multimode dynamics, non-Gaussian photon-number statistics, and negative intermode correlations.

The experimentally detected probe and conjugate fields correspond to coherent superpositions of the collective competition eigenmodes selected by the PM-fiber projection. Consequently, the measured Mandel-$Q$ parameter depends on the overlap between the detected spatial mode and the underlying collective modes, and may therefore vary with the fiber position, collection geometry, mode-field matching, and polarization alignment. The finite counting-bin duration determines the temporal correlations included in the measurement and therefore influences the measured photon-number statistics through temporal averaging, but it does not generate the mode competition itself.

Taken together, the combination of occupation-dependent gain saturation, multimode competition, spatial-mode projection by the PM fibers, and finite-time photon counting provides a physically consistent effective framework for understanding the experimentally observed sub-Poissonian photon-number statistics of the individual probe and conjugate fluxes.

\section*{Supplemental Note C: Dependence of the Mandel-$Q$ Parameter on Counting-Bin Width and Singles Count Rate}

The SFWM Mandel-$Q$ parameter was first evaluated for counting-bin widths ranging from $10$ to $1000~\mathrm{ns}$ using the analysis procedure described in Supplemental Note~D1. The measured Mandel-$Q$ parameters for the probe and conjugate fluxes are summarized in Table~\ref{tab:mandel_fwm_bin}. The Mandel-$Q$ parameter remains negative over the entire range of counting-bin widths and exhibits only a weak dependence on the bin width.

To verify that the observed negative Mandel-$Q$ parameters do not arise from detector-related effects or the data-analysis procedure, additional measurements were performed using an SPDC biphoton source at same single count rates. The measured SPDC Mandel-$Q$ parameters are summarized in Table~\ref{tab:q_biphoton} and remain positive over the entire range of counting-bin widths. 
\begin{table}[h]
\centering
\caption{Measured Mandel-$Q$ parameters for the SFWM probe and conjugate fluxes evaluated for different counting-bin widths. The uncertainties correspond to one standard deviation obtained from $10^{4}$ iterations.}
\label{tab:mandel_fwm_bin}
\small
\begin{tabular}{lcc}
\hline\hline
Bin width (ns) & $Q_{\mathrm{probe}}$ & $Q_{\mathrm{conjugate}}$ \\
\hline
10    & $-0.122 \pm 0.001$ & $-0.143 \pm 0.003$ \\
30    & $-0.631 \pm 0.002$ & $-0.673 \pm 0.002$ \\
50    & $-0.692 \pm 0.001$ & $-0.726 \pm 0.003$ \\
150   & $-0.711 \pm 0.002$ & $-0.732 \pm 0.003$ \\
250   & $-0.719 \pm 0.002$ & $-0.735 \pm 0.003$ \\
500   & $-0.734 \pm 0.002$ & $-0.754 \pm 0.004$ \\
1000  & $-0.749 \pm 0.003$ & $-0.762 \pm 0.004$ \\
\hline\hline
\end{tabular}
\end{table}

\begin{table}[h]
\centering
\caption{Measured Mandel-$Q$ parameters for a SPDC signal and idler fluxes evaluated for different counting-bin widths. The measurements were performed at singles count rates the same as those used in the SFWM measurements. The uncertainties correspond to one standard deviation obtained from $10^{4}$ iterations.}
\label{tab:q_biphoton}
\small
\begin{tabular}{lcc}
\hline\hline
Bin width (ns) &
$Q_{\mathrm{signal}}$ &
$Q_{\mathrm{idler}}$ \\
\hline
10   & $0.008 \pm 0.001$ & $0.012 \pm 0.001$ \\
30   & $0.022 \pm 0.003$ & $0.017 \pm 0.004$ \\
50   & $0.023 \pm 0.002$ & $0.030 \pm 0.003$ \\
100  & $0.027 \pm 0.003$ & $0.031 \pm 0.003$ \\
150  & $0.029 \pm 0.003$ & $0.032 \pm 0.003$ \\
250  & $0.030 \pm 0.002$ & $0.034 \pm 0.002$ \\
500  & $0.034 \pm 0.003$ & $0.038 \pm 0.002$ \\
1000 & $0.041 \pm 0.002$ & $0.042 \pm 0.003$ \\
\hline\hline
\end{tabular}
\end{table}


\begin{table}[h]
\centering
\caption{Measured Mandel-$Q$ parameters for the probe and conjugate fluxes at different singles count rates using a fixed counting-bin width of 150~ns.}
\label{tab:mandel_current_150ns_shifted}
\small
\begin{tabular}{lcc}
\hline\hline
Count rate ($s^{-1}$) & $Q_{\mathrm{probe}}$ & $Q_{\mathrm{conjugate}}$ \\
\hline
$4.91\times10^{6}$ & $-0.620$ & $-0.656$ \\
$5.30\times10^{6}$ & $-0.676$ & $-0.663$ \\
$5.90\times10^{6}$ & $-0.698$ & $-0.698$ \\
$6.20\times10^{6}$ & $-0.711$ & $-0.713$ \\
$6.53\times10^{6}$ & $-0.734$ & $-0.744$ \\
$6.93\times10^{6}$ & $-0.738$ & $-0.749$ \\
$7.68\times10^{6}$ & $-0.737$ & $-0.765$ \\
\hline\hline
\end{tabular}
\end{table}

To investigate the dependence of the measured photon-number statistics on the detected photon flux, additional measurements were performed over the experimentally accessible range of detected singles count rates. The Mandel-$Q$ parameter was evaluated using a fixed counting-bin width of 150~ns. The measured Mandel-$Q$ parameters for the probe and conjugate fluxes are summarized in Table~\ref{tab:mandel_current_150ns_shifted}. The Mandel-$Q$ parameter remains negative for both fluxes throughout the investigated photon-flux range, demonstrating that the observed sub-Poissonian photon-number statistics persist over the experimentally accessible range of detected photon fluxes. At lower pump powers, the nonlinear SFWM contribution would be increasingly obscured by the fluorescence background associated with linear absorption, thereby limiting the accessible low-flux range of the measurement. At higher pump powers, the generated photon flux exceeds the linear operating range of the single-photon detectors, thereby setting the upper limit of the accessible photon-flux range.

\section*{Supplemental Note D: Data analysis}

The data-analysis procedures described below were used for the coherent, SPDC biphoton, and SFWM measurements. All analyses were performed using the recorded time-tagged photon-arrival data acquired simultaneously from the probe and conjugate using single-photon counting modules (SPCMs). The procedures used to evaluate the Mandel-$Q$ parameter and the temporal-correlation measurements are described below.

\subsection*{D1. Mandel-$Q$ analysis}

The Mandel-$Q$ parameter was evaluated directly from the recorded time-tagged photon-arrival data. The analysis was performed independently for the probe and conjugate detection channels. For a selected counting-bin width $T_{\mathrm{bin}}$, which defines the photon-counting integration window, the complete acquisition time was divided into consecutive non-overlapping intervals of equal duration. The number of singles counts within each interval was recorded, yielding the photon-number sequence $\{n_k\}$.

The mean photon number and the corresponding variance were calculated as
\[
\langle n\rangle
=
\frac{1}{N}
\sum_{k=1}^{N}
n_k,
\]
and
\[
\mathrm{Var}(n)
=
\frac{1}{N-1}
\sum_{k=1}^{N}
\left(
n_k-\langle n\rangle
\right)^2,
\]
respectively, where $N$ denotes the total number of counting intervals. The Mandel-$Q$ parameter was then evaluated from
\[
Q
=
\frac{\mathrm{Var}(n)-\langle n\rangle}
{\langle n\rangle}.
\]
This procedure was repeated for different counting-bin widths to investigate the dependence of the measured photon-number statistics on the integration time. The minimum counting-bin width was intentionally chosen to be shorter than the detector dead time to reveal the influence of dead-time-induced artifacts on the measured photon-number statistics.

\subsection*{D2. Temporal-correlation analysis}

The temporal-correlation measurements were obtained from the same time-tagged photon-arrival data used for the Mandel-$Q$ analysis. For each photon detected in the conjugate channel, the probe detection event having the smallest absolute arrival-time difference,
\[
|t_{\mathrm{probe}}-t_{\mathrm{conjugate}}|,
\]
was identified from the recorded timestamps. The corresponding arrival-time delay was then calculated as
\[
\tau
=
t_{\mathrm{probe}}
-
t_{\mathrm{conjugate}},
\]
where $t_{\mathrm{probe}}$ denotes the arrival time of the selected probe photon. Consequently, each conjugate photon contributes a single arrival-time delay to the coincidence histogram, providing a unique temporal delay for every detected event without introducing a predefined coincidence window.

The calculated arrival-time delays were accumulated into a coincidence histogram using a uniform bin width of 250~ps over the interval$-30~\mathrm{ns}
\le
\tau
\le
30~\mathrm{ns},$ yielding the experimental temporal-correlation profile presented in the main text. The experimental coincidence histogram was subsequently fitted using the theoretical model described in Supplemental Note~A, which includes the dominant SFWM contribution together with the two additional Raman-assisted scattering channels.

\section*{Supplemental Note E: Calibration Procedure for free-space and fiber-coupled squeezing measurements}

The intensity-difference squeezing measurements presented in the main text were performed using two different balanced-detector configurations. The free-space measurements used a modified Thorlabs PDA450A balanced detector equipped with Hamamatsu photodiodes to improve the detection efficiency at 795~nm. Owing to the experimental constraints of the fiber-coupled configuration, this detector could not be used after coupling the twin beams into the PM fibers. Consequently, the fiber-coupled measurements were performed using a standard Thorlabs PDA450A balanced detector.

To quantify the relative difference between the two detector configurations, both detectors were characterized under identical free-space measurement conditions using the same optical power and alignment. Under these conditions, the standard detector consistently measured approximately $1~\mathrm{dB}$ less intensity-difference squeezing than the modified detector. This difference arises from the lower overall detection efficiency of the standard detector and was found to be essentially independent of the analysis frequency over the measurement bandwidth.

Since the measured difference between the two detector configurations is approximately $1~\mathrm{dB}$ and is essentially independent of the analysis frequency, a constant $1~\mathrm{dB}$ correction was applied to the entire fiber-coupled squeezing spectrum to enable a direct comparison with the free-space measurement. This correction produces a uniform vertical shift of the entire spectrum without altering its spectral shape or frequency dependence. Consequently, the free-space and fiber-coupled spectra exhibit an apparent crossover at the highest analysis frequencies. This crossover should not be interpreted as a physical improvement in squeezing after propagation through the polarization-maintaining fibers. Rather, it is a consequence of the relative detector calibration used to compensate for the different detection efficiencies of the two balanced-detector configurations. The calibration places both measurements on the same detection-efficiency scale, allowing a direct comparison of their squeezing performance while preserving the intrinsic frequency dependence of the measured spectra.



\end{document}